\documentclass[aps,prb,reprint,superscriptaddress,floatfix]{revtex4-2}
\usepackage[T1]{fontenc}
\usepackage{amsmath}
\usepackage{booktabs}
\usepackage{glossaries}
\usepackage{graphicx}
\usepackage{hyperref}
\usepackage{enumitem}
\usepackage{multirow}
\usepackage[table]{xcolor}
\usepackage{cleveref}
\usepackage{mhchem}
\usepackage{siunitx}
\usepackage{makecell}

\newacronym[plural=APIs, firstplural=application programming interface (APIs)]{api}{API}{application programming interface}
\newacronym{cli}{CLI}{command line interface}
\newacronym{url}{URL}{Uniform Resource Locator}
\newacronym{cod}{COD}{Crystallographic Open Database}
\newacronym{icsd}{ICSD}{Inorganic Crystal Structure Database}
\newacronym{mpds}{MPDS}{Materials Platform for Data Science}
\newacronym[plural=CIFs, firstplural=Crystallographic Information Files (CIFs)]{cif}{CIF}{Crystallographic Information File}
\newacronym{sssp}{SSSP}{Standard Solid State Pseudopotentials}
\newacronym{scf}{SCF}{self-consistent field}
\newacronym{pbe}{PBE}{Perdew-Burke-Ernzerhof}
\newacronym{pbesol}{PBEsol}{Perdew-Burke-Ernzerhof functional revised for solids}
\newacronym{qe}{QE}{\textsc{Quantum ESPRESSO}}
\newacronym{gpu}{GPU}{Graphics Processing Unit}
\newacronym{rsw}{RSW}{reconnaissance SCF workflow}
\newacronym{gga}{GGA}{Generalized Gradient Approximation}
\newacronym{dft}{DFT}{Density-functional theory}
\newacronym{eh}{EH}{error handler}
\newacronym{mc3d}{MC3D}{Materials Cloud Three-Dimensional Structure Database}
\newacronym{mc3ds}{MC3D-source}{MC3D-source}
\newacronym{mp}{MP}{Materials Project}
\newacronym{oqmd}{OQMD}{Open Quantum Materials Database}
\newacronym{3d}{3D}{three-dimensional}

\definecolor{DARKMAGENTA}{HTML}{af2f40}
\definecolor{darkgreen}{HTML}{00aa00}

\hypersetup{
    colorlinks=true,
    linkcolor=DARKMAGENTA,
    citecolor=DARKMAGENTA,
    urlcolor=DARKMAGENTA,
}

\newcommand\code[1]{\texttt{#1}}
\newcommand\ccite[1]{~\cite{#1}}

\begin{document}

\title{MC3D: The Materials Cloud computational database of experimentally known stoichiometric inorganics}

\author{Sebastiaan P. Huber$^*$}
\affiliation{Theory and Simulation of Materials (THEOS), and National Centre for Computational Design and Discovery of Novel Materials (MARVEL), \'Ecole Polytechnique F\'ed\'erale de Lausanne (EPFL), 1015 Lausanne, Switzerland}

\author{Michail Minotakis$^*$}
\affiliation{PSI Center for Scientific Computing, Theory, and Data, 5232 Villigen PSI, Switzerland}

\author{Marnik Bercx$^*$}
\affiliation{Theory and Simulation of Materials (THEOS), and National Centre for Computational Design and Discovery of Novel Materials (MARVEL), \'Ecole Polytechnique F\'ed\'erale de Lausanne (EPFL), 1015 Lausanne, Switzerland}
\affiliation{PSI Center for Scientific Computing, Theory, and Data, 5232 Villigen PSI, Switzerland}

\author{Timo Reents}
\affiliation{PSI Center for Scientific Computing, Theory, and Data, 5232 Villigen PSI, Switzerland}

\author{Kristjan Eimre}
\affiliation{Theory and Simulation of Materials (THEOS), and National Centre for Computational Design and Discovery of Novel Materials (MARVEL), \'Ecole Polytechnique F\'ed\'erale de Lausanne (EPFL), 1015 Lausanne, Switzerland}
\affiliation{PSI Center for Scientific Computing, Theory, and Data, 5232 Villigen PSI, Switzerland}

\author{Nataliya Paulish}
\affiliation{PSI Center for Scientific Computing, Theory, and Data, 5232 Villigen PSI, Switzerland}

\author{Nicolas H\"ormann}
\affiliation{Fritz-Haber-Institut der Max-Planck-Gesellschaft, 14195 Berlin, Germany}

\author{Martin Uhrin}
\affiliation{Université Grenoble Alpes, MIAI Cluster IA, SIMaP, 38000 Grenoble, France}

\author{Nicola Marzari}
\affiliation{Theory and Simulation of Materials (THEOS), and National Centre for Computational Design and Discovery of Novel Materials (MARVEL), \'Ecole Polytechnique F\'ed\'erale de Lausanne (EPFL), 1015 Lausanne, Switzerland}
\affiliation{PSI Center for Scientific Computing, Theory, and Data, 5232 Villigen PSI, Switzerland}
\affiliation{Bremen Center for Computational Materials Science, and MAPEX Center for Materials and Processes, University of Bremen, 28359 Bremen, Germany}

\author{Giovanni Pizzi}
\affiliation{Theory and Simulation of Materials (THEOS), and National Centre for Computational Design and Discovery of Novel Materials (MARVEL), \'Ecole Polytechnique F\'ed\'erale de Lausanne (EPFL), 1015 Lausanne, Switzerland}
\affiliation{PSI Center for Scientific Computing, Theory, and Data, 5232 Villigen PSI, Switzerland}

\date{\today}

\begin{abstract}
\gls{dft} is a widely used method to compute properties of materials, which are often collected in databases and serve as valuable starting points for further studies.
In this article, we present the \gls{mc3d}, an online database of computed \gls{3d} inorganic crystal structures.
Close to a million experimentally reported structures were imported from the COD, ICSD and MPDS databases; these were parsed and filtered to yield a collection of $72\,589$ unique and stoichiometric structures, of which 95\% are, to date, classified as experimentally known.
The geometries of structures with up to 64 atoms were then optimized using \gls{dft} with automated workflows and curated input protocols.
The procedure was repeated for different functionals (and computational protocols), with the latest version (\gls{mc3d} PBEsol-v2) comprising $32\,013$ unique structures.
All versions of the \gls{mc3d} are made available on the Materials Cloud portal, which provides a graphical interface to explore and download the data.
The database includes the full provenance graph of all the calculations driven by the automated workflows, thus establishing full reproducibility of the results and more-than-FAIR procedures.
\end{abstract}

\keywords{high-throughput, materials database, scientific workflow, automation, provenance, reproducibility}

\maketitle

\section*{Introduction}
The paradigm of computational materials discovery based on quantum-mechanical approaches has found significant adoption in recent years\cite{Oganov:2019,Marzari:2021}.
Computational studies have grown in number and scale, in particular those based on \gls{dft}\cite{Hohenberg:1964,Kohn:1965}, a powerful first-principles method to compute the electronic ground state of materials, and routinely used to predict their properties\cite{VanNoorden:2014,VanNoorden:2025}.
The predictive power of \gls{dft}, together with the growing availability of powerful computational resources, has driven high-throughput computational materials discovery~\cite{Curtarolo:2013}, aiming to screen or discover materials with optimal properties.
This has in turn spurred the development of several workflow systems to manage the large amount of calculations and the data they produce\cite{Huber:SciData:2020, Uhrin:CompMatSci:2021,Rosen:2024,Jain:2015,HjorthLarsen:2017,Janssen:2019,Calderon:2015, Oses:2023,Adorf:2018,Rosen:2025,Cunningham:2023,Choudhary:2019,Speckhard:2025}, leading to the creation of multiple publicly available databases of computed materials properties\cite{Landis:2012,Curtarolo:2012,Jain:2013,Saal:2013,Kirklin:2015,Draxl:2018,Choudhary:2020,Wang:2022,Schmidt:2022,Shen:2022,barroso-luque:2024,kaplan:2025}.
Recently, these databases have been a catalyst for the adoption of machine-learning methods in computational materials science, as the basis to train models\cite{batatia:2024,yang:2024,Mazitov:2025, rhodes:2025,fu:2025}, to predict the properties of known materials\cite{Levamaki:2022,Stanev:2018,Isayev:2017,Pilania:2016, pota:2024,Loew:2025} or even the existence of new ones\cite{Mazhnik:2020,Avery:2019,Gu:2022,Schmidt:2021,zeni:2025,xie:2022}.
To ease the ingestion of data from these databases, a community effort has resulted in the OPTIMADE universal API~\cite{Andersen:2021,Evans:2024} that enable users of the participating databases to use a common query language and obtain results in a common format.

Nevertheless, computational databases might not always use a consistent setup to compute the properties of all materials.
This, in turn, leads to potential small inconsistencies in the data across the periodic table, making more challenging to obtain accurate machine-learning models to predict properties of materials not present in the database, as already noted in Ref.~\onlinecite{Dragoni:2018} and recently demonstrated by the accuracy of the PET-MAD model~\cite{Mazitov:2025,Mazitov:2025a}.

In this article, we first present a set of automated workflows that import crystal structures from the \gls{cod}\ccite{web:cod}, the \gls{icsd}\ccite{web:icsd}, and the \gls{mpds}\ccite{web:mpds}.
Using these workflows, we import all entries (close to a million structures, mainly reported from experiments) and reduce them to a set of $72\,589$ unique bulk stoichiometric inorganic crystal structures, after several filtering steps that we describe in detail below.
For the subset of structures not including lanthanides or actinides, and including at least all systems with up to 64 atoms, we compute the electronic ground state of their optimized geometries via \gls{dft} using the open-source \gls{qe}\ccite{Giannozzi:2009,Giannozzi:2017} code, powered by the SIRIUS library\ccite{web:sirius}.
The computational inputs for the \gls{dft} simulations executed by the workflows are determined by fully automated protocols, requiring only the input structure as mandatory input, therefore enabling automated high-throughput execution of the workflows.
The protocols are tuned to select input parameters that optimize the balance between precision and computational cost.
Since the workflows are implemented in AiiDA\ccite{Pizzi:2016,Huber:SciData:2020,Uhrin:CompMatSci:2021}, the entire provenance of each optimized geometry, including all raw simulation inputs and outputs, is automatically preserved in the database.

The calculations use both PBE\ccite{Perdew:1996} and PBEsol\ccite{Perdew:2008}, and different protocols for the input parameters.
We refer to the collection of resulting databases of optimized geometries as the \glsentrylong{mc3d} (\glsentryshort{mc3d}). 
In particular, when considering the PBEsol functional and the most recent protocol \texttt{PBEsol-v2}, we provide $32\,013$ unique structures computed and relaxed with \gls{dft}.

We make the \gls{mc3d} database fully and publicly available through the Materials Cloud Archive\cite{web:Huber:MCA:2025:mc3d}.
Moreover, for easier accessibility, we provide a dedicated web application in the Materials Cloud\cite{Talirz:2020} platform (\href{https://www.materialscloud.org/mc3d}{https://www.materialscloud.org/mc3d}), through which all data can be inspected interactively, as well as an OPTIMADE-compliant endpoint to access the data via a standardized API.
In the following, we describe the data import and processing pipeline, the computational workflows, the content of the \gls{mc3d}, as well as its web interface.

\section*{Results}
Our pipeline started from importing $901\,210$ crystal structures from three of the largest experimental inorganic crystal structure databases: the \gls{cod}\ccite{web:cod} (revision 213553), the \gls{icsd}\ccite{web:icsd} (version 2017.2), and the \gls{mpds}\ccite{web:mpds} (version 1.0.2).

Fig.~1(a) shows a graphical representation of the pipeline that reduced this initial collection to $72\,589$ unique stoichiometric crystal structures.
In the first step of the pipeline, the input structures---obtained in the \gls{cif} format\ccite{Hall:1991}---were parsed and validated.
A number of \gls{cif} files had to be discarded due to invalid syntax or inconsistent information.
From the $723\,165$ valid \glspl{cif}, the crystal structure was successfully parsed and the crystal lattice was normalized reducing it to the primitive cell (see the Methods for details on the validation and parsing).

\begin{figure*}[t]
    \centering
    \includegraphics[width=16cm]{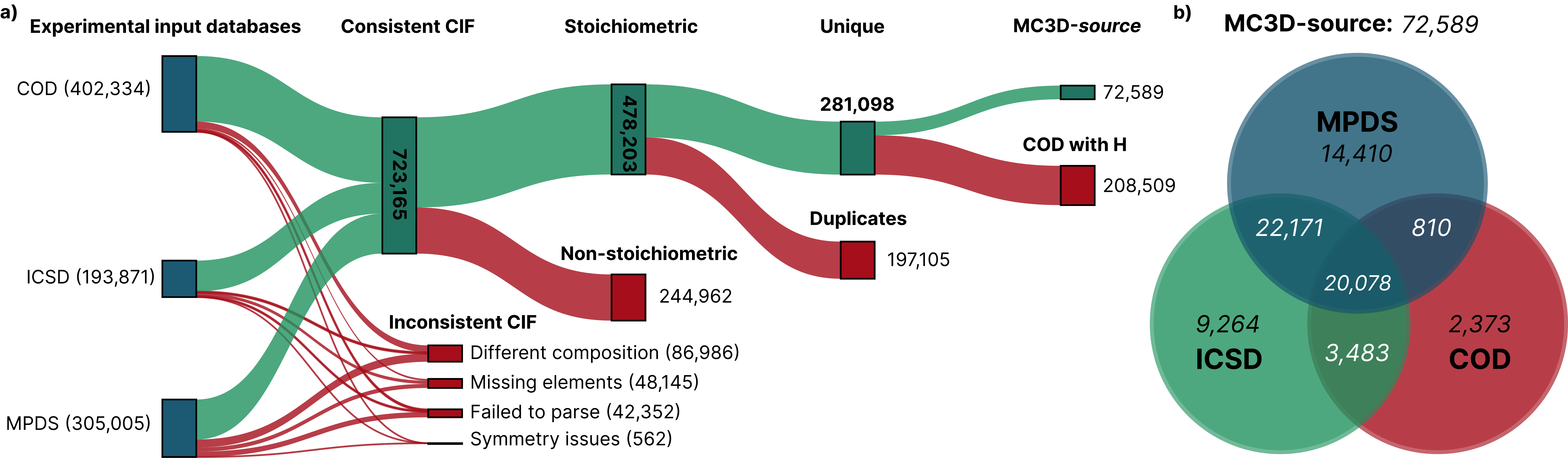}
    \caption{
        (a) Sankey diagram visualizing the pipeline that filtered the $901\,210$ \gls{cif} files, as imported from the COD, ICSD, and MPDS databases, down to the \gls{mc3ds} collection of $72\,589$ unique stoichiometric inorganic crystal structures.
        Red branches indicate structures that were discarded, while green branches correspond to structures that made their way into the following filtering step.
        The first step discarded \gls{cif} files that contained invalid syntax or inconsistent information, or could otherwise not be parsed to yield a valid crystal structure definition.
        Non-stoichiometric structures are discarded in the second step.
        Out of the remaining structures, $197\,105$ were determined to be duplicates in the third step. In the last step, $208\,509$ structures only available in COD and including hydrogen atoms were not considered for further analysis because they typically correspond to molecular crystals.
        (b) Venn diagram of the distribution of unique structures in the \gls{mc3ds} collection from the three source databases.
    }
\end{figure*}

The next step in the pipeline filtered out $244\,962$ structures that are not stoichiometric, i.e., those that contain partial occupancies.
These structures are beyond the scope of the current study, due to the requirement of considering appropriately chosen large supercells to accommodate partial occupancies.
From the remaining $478\,203$ crystal structures, $197\,105$ were found to be duplicates (details on how uniqueness is determined are given in the Methods section).
The last filtering step in the pipeline removed $208\,509$ structures that include hydrogen and are available only in the \gls{cod} database.
As detailed in the Methods section, we apply this filter to effectively exclude molecular crystals, included in \gls{cod} but not in \gls{icsd} nor in \gls{mpds}, since our study focuses on inorganic compounds.

The final collection of starting crystal structures, labeled \gls{mc3ds}, consists of $72\,589$ unique three-dimensional crystal structures.
Of these, $3\,305$ structures are explicitly reported to have a theoretical origin by the source database.
Therefore, we identify no more than $69\,284$ experimentally known stoichiometric inorganics (see Methods for more details), a much smaller number than typically quoted.
This first result already highlights the relatively small size of the space of experimentally known inorganic stoichiometric crystal structures.
This might point to the actual size of this space, as well as the complexity of synthesizing and characterizing complex systems, especially those including several chemical elements (ternaries, quaternaries and beyond), as already highlighted in Ref.~\onlinecite{Davies:2016}.

Fig.~1(b) shows the Venn diagram of the distribution of structures in the \gls{mc3ds} from the three source databases.
Out of the total $72\,589$ structures, $20\,078$ structures are present in all three databases, and $46\,542$ structures are present in at least two of the three databases.
The \gls{cod}, \gls{icsd} and \gls{mpds} each contain $2\,373$, $9\,264$ and $14\,410$ structures that are unique to them, respectively (or $210\,882$ for \gls{cod}, if we consider also structures with hydrogen, excluded from \gls{mc3ds} in our last filtering step as discussed above).

\begin{figure*}[t]
    \centering
    \includegraphics[width=16cm]{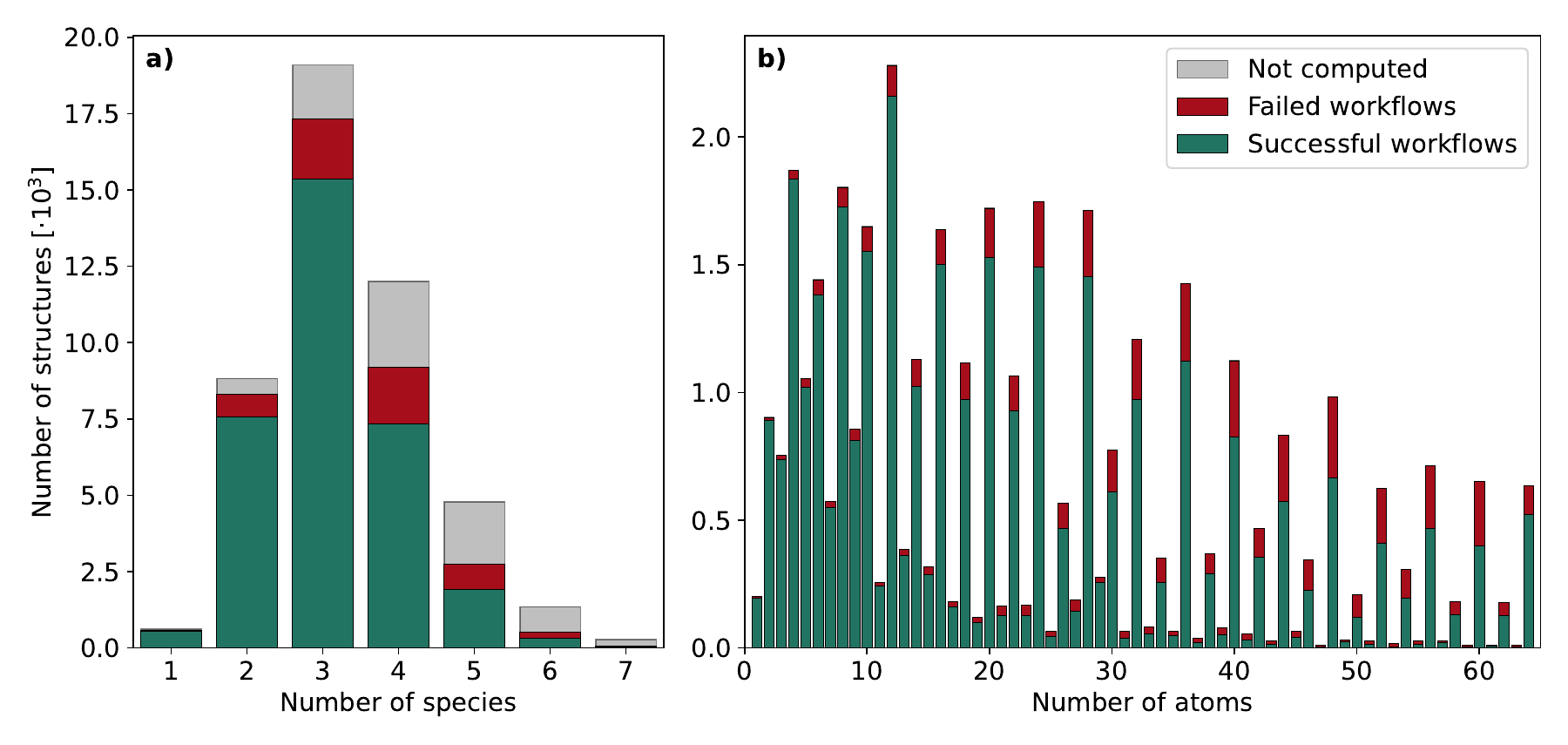}
    \caption{
        Histograms of the results of the geometry-optimization workflows executed on the subset of \gls{mc3ds} excluding lanthanides and actinides ($46\,964$ structures) as a function of (a) the number of species (i.e., distinct chemical elements) in the structure, and (b) the number of atoms in the primitive cell.
        Green represents successful workflows, whereas red indicates workflows that failed with an unrecoverable error.
        The gray bars in (a) correspond to structures with more than 64 atoms that have not been computed.
        In panel (a), the 34 structures in \gls{mc3ds} with 8 or more species are not shown in the plot.
    }
\end{figure*}

A subset of the $72\,589$ unique crystal structures of the \gls{mc3ds} was then passed through our automated workflow to optimize their geometry using \gls{dft}.
First, structures containing lanthanides or actinides were not considered.
This is because, on one hand, a physically accurate description of their electronic structure (in particular of the $f$ electrons) would require a more sophisticated treatment than standard DFT.
On the other hand, the pseudopotentials currently available for these elements, when available, have limited precision (see, e.g., Fig.~S9.7 in the Supplementary Information of Ref.~\onlinecite{Bosoni2024}) and also often lead to numerical instabilities and a significant failure rate.
The subset of the \gls{mc3ds} database that excludes lanthanides and actinides contains $46\,964$ structures.
Furthermore, in order to better exploit the available computational budget, we prioritized processing structures with smaller number of atoms in the unit cell.
While also some larger structures have already been computed, in this paper we focus the analysis on the structures with up to $64$ atoms, all of which were submitted to the ground-state atomic and geometric relaxation workflow.

We considered two different \gls{dft} functionals (PBE and PBEsol) and, in the PBEsol case, we run the workflow with two different versions of the input parameter protocols, the second one (v2) being refined and whose numerical precision was verified via extensive tests\cite{Nascimento:2025}.
The resulting three databases of optimized geometries are versions of the \gls{mc3d} and are referred to as \texttt{PBE-v1}, \texttt{PBEsol-v1} and \texttt{PBEsol-v2}, respectively (see Methods for an overview of the differences between the different input parameters and pseudopotential libraries for the \gls{mc3d} versions).
The rest of this article focuses on the results of the most accurate and precise version, \texttt{PBEsol-v2}, but the results of all three versions are publicly available.

Out of the $38\,739$ structures that have been processed, the relaxation workflow successfully completed $33\,142$ structures, corresponding to a success rate of $85.5\%$.
In the remaining cases, the workflow encountered errors that could not be solved by the automatic error handling mechanisms implemented (see the Section on automated error handling in the Methods for more details).
These results are represented visually in Fig.~2, where the workflow results are plotted as a histogram as a function of (a) the number of elemental species and (b) the number of atomic sites in the structure.

\begin{figure}[t]
    \centering
    \includegraphics[width=8cm]{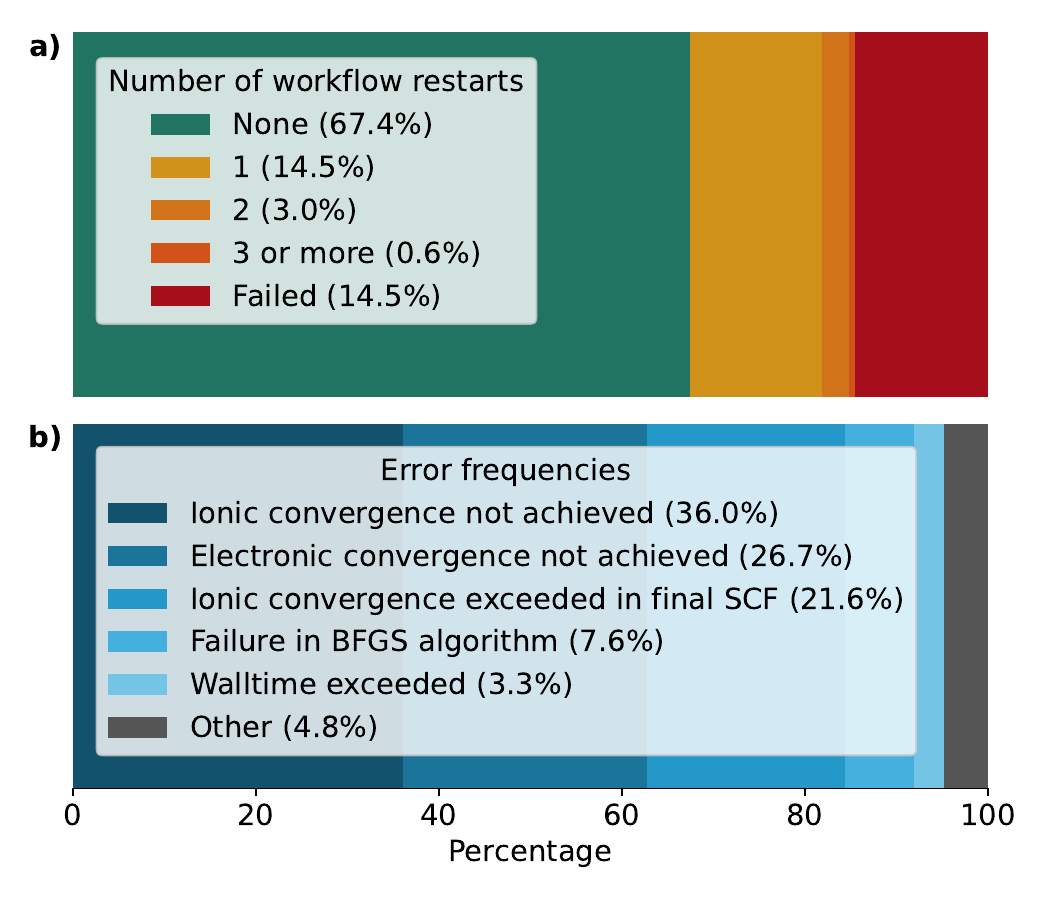}
    \caption{
        Bar charts of (a) the number of restarts for each geometry optimization workflow and (b) the frequency of the 5 most common observed errors (including those that were recovered).
        The vast majority of workflows finish after the first iteration ($67.4\%$).
        Workflows that need 1, 2, or 3 or more restarts are less common with $14.5\%$, $3.0\%$, and $0.6\%$, respectively.
        About $14.5\%$ of workflows fail to complete successfully despite the error handling mechanism.
        The most common observed errors are the failure of the \texttt{pw.x} simulation code to achieve ionic or electronic convergence within the maximum allowed number of iterations.
    }
\end{figure}

Our complete relaxation workflow to compute the optimized geometry of a crystal structure and its ground-state electronic charge density comprises multiple steps executing the \code{pw.x} code of the SIRIUS-accelerated\cite{web:sirius} \gls{qe} package.
However, the \texttt{pw.x} code can encounter a variety of errors, ranging from issues with the compute hardware on which the code is run itself (e.g., insufficient job resources or node failures) to errors during code execution (e.g., numerical instabilities or convergence issues).
These errors need to be handled as much as possible automatically by the workflow in order to be robust and scalable.
Fig.~3 shows a visual representation of how often a workflow needed to restart the calculation to achieve successful completion, and of the most commonly observed errors.
We first highlight that over $67\%$ of the workflows completed successfully after the first execution of \texttt{pw.x} (green bar in Fig.~3a) without triggering any error handling.
We stress that this success rate could be achieved also thanks to the appropriate choices of input parameters defined by our protocols, that prevent certain failure modes, and by the use of advanced direct minimization algorithms for electronic convergence beyond the standard iterative self-consistent field (SCF) approach, such as direct-minimization strategies~\cite{Marzari:1997,Freysoldt:2009} as implemented in the \texttt{nlcglib}\cite{web:nlcglib} plugin of SIRIUS library~\cite{web:sirius}.
The vast majority of errors come from the failure to converge to the required precision either the geometry optimization or the self-consistent field calculation (within the maximum allowed number of iterations).
Notably, the error handling mechanism implemented in our workflows (see the Methods section for more details) manages to recover from these errors in over half of the cases in which a single \texttt{pw.x} execution fails (orange bars in Fig.~3a, with one or more workflow restarts).
We emphasize that the error handler can decide to change the input of the calculation before resubmission, depending on the type of error reported in the code output. However, such changes do not include physical parameters of the calculation that would affect the results, but are limited only to those numerical aspects that can increase the chance of convergence (e.g., reducing the mixing parameters of the self-consistent cycle or changing diagonalization algorithm).
Nevertheless, despite our error-handling mechanisms, for a number of crystal structures ($14.5\%$ of the total, corresponding to the rightmost red segment in Fig.~3) the workflow failed to complete after exhausting the configured maximum number of retries.
Work on improving the error handling and robustness of the workflows is ongoing, and we expect to further increase the success rate in future versions of the \gls{mc3d}.
We also stress that, because of the nature of the errors, achieving a significant increase in convergence rate cannot be obtained solely by improving workflow error-handling mechanisms, but also requires enhancements to the underlying algorithms in the \texttt{pw.x} code, as already addressed in this work by using the advanced convergence algorithms implemented in SIRIUS~\cite{web:sirius} discussed above.

\begin{figure*}[t]
    \centering
    \includegraphics[width=16cm]{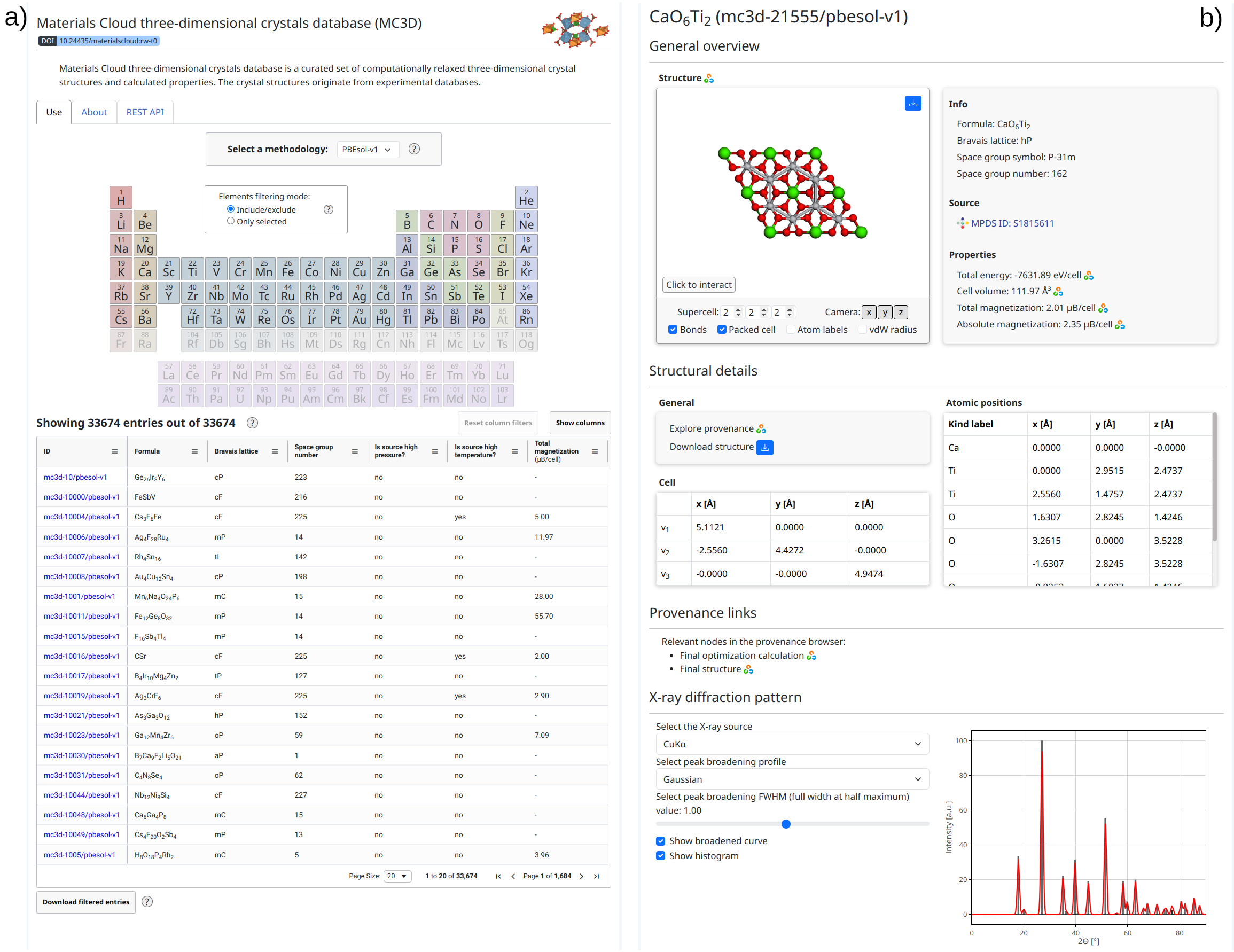}
    \caption{
        Screenshot of the web application, integrated into the Materials Cloud, that provides a visual interface to explore the structures of the \gls{mc3d} and the associated computed properties.
        (a) The landing page of the web application, providing an intuitive interface to select the desired version of the \gls{mc3d} and a periodic table to filter for materials based on their composition.
        Moreover, the index table below gives an overview of the filtered structures together with a number of structural properties, and a link to a detail page.
        (b) The detail page for a specific material, showing detailed information and various computed properties, as well as links to the source calculation in the AiiDA provenance graph, providing direct access to the full provenance of how the property was obtained.
    }
\end{figure*}

To make the \gls{mc3d} easily accessible and explorable, we make it available as a Discover section on the Materials Cloud\ccite{Talirz:2020} portal at \url{https://www.materialscloud.org/mc3d}.
Fig.~4 shows screenshots of the web application.
The landing page allows the user to select a version of the \gls{mc3d} and filter structures based on their composition via a periodic table.
The matched structures are displayed in an index table that provides a link to a detail page together with several structural and material properties.
The user can interactively edit the list of visible columns, adjust the row sorting, and apply column-based filtering.
The detail pages give a detailed overview of the structural properties of the crystal as well as any other computed property.
In particular, we provide powder X-ray diffraction (XRD) patterns computed using \texttt{pymatgen}~\cite{Ong:2013} based on Bragg's law and atomic scattering factors. Reflections can be plotted using Gaussian or Lorentzian peaks with an area corresponding to peak intensity, and the full width at half maximum (FWHM) parameter can be set by the user.
Notably, the full data and calculation provenance is directly accessible from the web interface.
Computed properties are decorated with an AiiDA icon linking to the source calculation in the Materials Cloud Explore section.
There, users can browse the full provenance graph of MC3D, extract raw inputs and outputs, and reconstruct how the optimized geometries were obtained.
A more detailed description of the web platform is provided in the Methods section.

\section*{Discussion}

\begin{figure}[t]
    \centering
    \includegraphics[width=8cm]{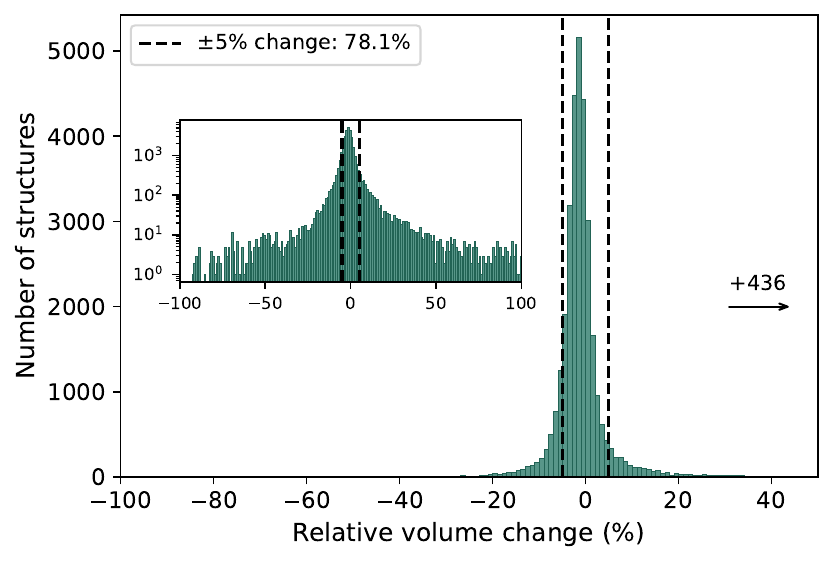}
    \caption{
        Histogram of the relative differences between the cell volumes of the input and optimized structures for the \texttt{PBEsol-v2} MC3D database. The bin size is $1\%$.
        A positive (negative) difference indicates that the optimized structure expanded (compressed) compared to the input structure.
        The majority of structures ($78.1\%$) falls within the range of $-5\%$ to $5\%$, as indicated by the vertical dashed black lines.
        There are 436 structures that have a relative volume change greater than $50\%$ and are not shown in the image.
    }
\end{figure}

Our main goal is to provide a curated set of crystal structures, optimized with \gls{dft} using a consistent protocol, focusing on experimentally known structures.
In this way, if structures from \gls{mc3d} are used as a starting point for further screening studies, optimal candidates are already available for additional experiments.
We highlight however that, while the source databases typically focus on experimental structures, they also report theoretical structures in some cases.
By inspecting the metadata reported by the source databases, we could flag $2\,586$ structures in \gls{mc3d} \texttt{PBEsol-v2} that originate from structures tagged as theoretical (see details in the Methods section).
This information can be accessed by the corresponding column in the Materials Cloud web interface.
We stress that, however, we need to rely on the information provided by the source databases which might not always be accurate or complete.
Therefore, users should be aware that, in some rare cases, some of the structures in the \gls{mc3d} might not be experimentally known, even if they are not flagged as theoretical.

For the $33\,142$ successfully completed workflows, the optimized geometries were validated by comparing the difference in volume between the initial and optimized crystal structures, as large volume changes could be an indication of problems with either the initial or optimized structure.
Fig.~5 shows a histogram of the relative volume change in percent between the initial and optimized geometry.
The majority of optimized geometries have a change in volume of $\pm 5\%$, indicating that the optimized geometry with \gls{dft} is not too far away from the experimentally determined source crystal structure.

The changes in volume between the source and optimized geometry are due to various factors.
First of all, the source databases include thousands of layered structures~\cite{Mounet:2018,Campi:2023}.
However, since in this work we are not including van der Waals corrections in the \gls{dft} calculations, the optimized geometries of these layered structures will typically have a potentially much larger cell volume than the source structure~\cite{Mounet:2018}.
Moreover, some source structures may have been characterized at high pressure and/or high temperature conditions (whereas our computational methods are limited to a temperature of 0~K and we considered a target zero pressure), which can also highly influence cell volume.

To account for these factors, we extracted the metadata from the source databases regarding the experimental conditions of the source structures, i.e., the temperature and pressure.
Unfortunately, the conditions at which structures have been characterized are only very rarely accurately reported in the \gls{cif} source, making it difficult to reliably classify the structures.
For those where such data is available, we classify them as high pressure (high temperature) if the pressure (temperature) is reported and above 5000~atm (100~$^\circ$C).
In Fig.~6(a) we show a histogram of the relative volume change for structures classified as above, only including those structures that report explicitly both pressure and temperature, highlighting that the vast majority of the outliers are structures that are either high pressure or high temperature.

However, this plot includes very few structures, and in particular only very few (21) explicitly report ``ambient'' conditions (according to our definition of both pressure and temperature below 5000~atm and 100~$^\circ$C, respectively).
We therefore also plot in Fig.~6(b) the distribution of implicitly defined ambient structures, i.e., those that only report one of the two conditions (pressure or temperature), and that condition is below our thresholds (i.e., we assume that the other condition is also ambient).
Although the majority of the most extreme outliers are still high pressure or high temperature, there are more ambient structures whose volume changes significantly.
To better understand these outliers, we investigated the source papers of the 7 explicit ambient outliers outside the $[-5\%, 5\%]$ range, as well as the 43 implicit ambient outliers outside the $[-50\%, 50\%]$ range.
Among the explicit ambient structures, we found that 6 were actually extracted from a high pressure/temperature study (but this was not correctly reported in the source databases), and one was a layered structure.
Among the implicit ambient structures, also in this case about half were part of high pressure/temperature studies, layered structures, or molecular crystals.
For other structures, the atomic occupations were most likely not fully stoichiometric, even if reported as such in the CIF file.
Therefore, although we cannot explain all ambient outliers, it is clear that many of them are related to a possible incorrect reporting of the conditions or of the crystal structure, thus highlighting the importance of careful curation of the source databases.

\begin{figure}[tb]
    \centering
    \includegraphics[width=8cm]{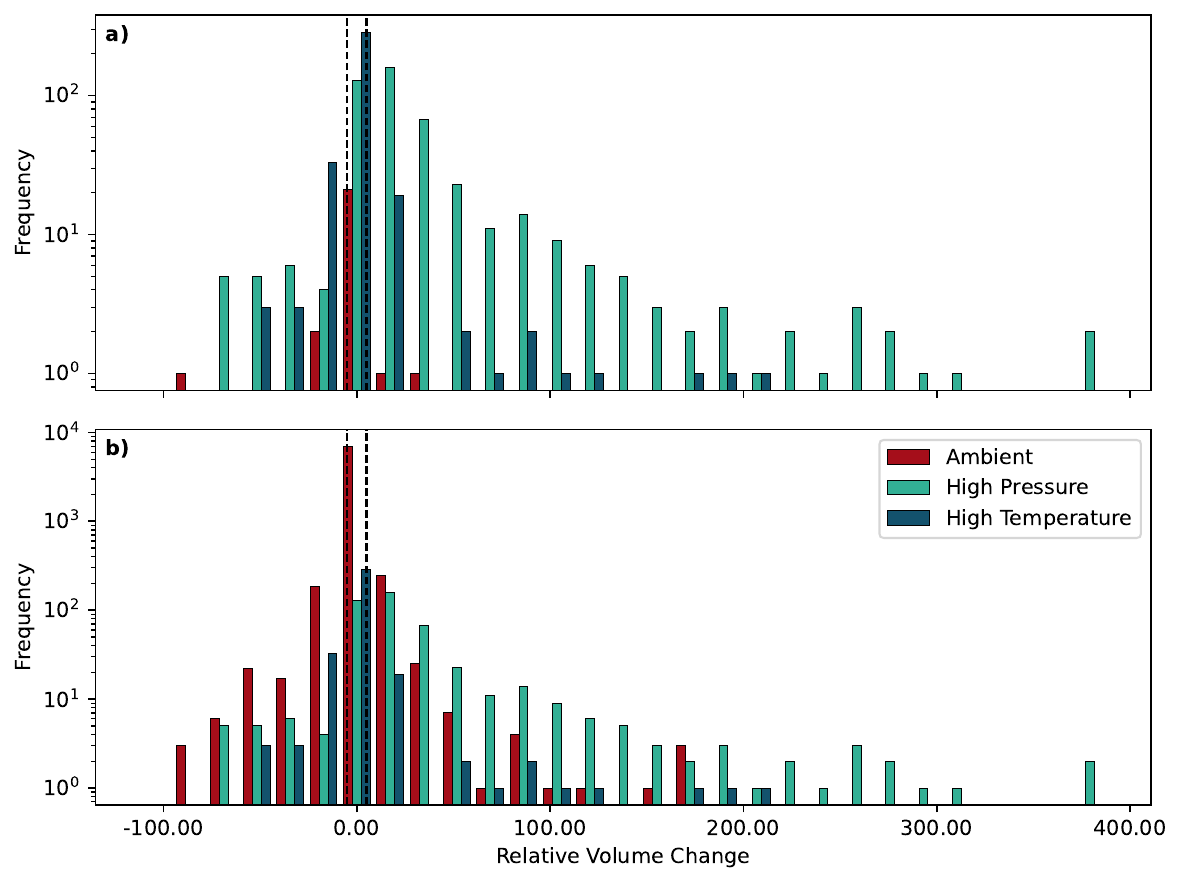}
    \caption{
    Histogram of the relative volume change of structures, sorted into the three categories "Ambient", "High Pressure" and "High Temperature". (a) Structures with explicit ambient conditions: only structures where both pressure and temperature are reported in the source database are included in the plot.
    See main text for the thresholds defining high pressure and temperature.
    Structures that are both high pressure and high temperature are counted in both categories.
    (b) Structures with implicit ambient conditions: also structures where only one between pressure and temperature is reported in the source database, and this condition is below our thresholds, are included in the plot (and included in the ``Ambient'' group).
    }
    \label{fig:non-ambient}
\end{figure}

\begin{table*}[ht]
    \centering
    \begin{tabular*}{\linewidth}{l@{\extracolsep{\fill}} c c c c}
    \toprule
    New in \gls{mc3d}: & Subtype A & Subtype B & Subtype C & Total new\\
    Reference & & & & \\
    \midrule
    Materials Project\footnote{As downloaded using the Materials Project REST API in May 2025.} & $2\,647$ & $2\,704$ & $1\,139$ & $6\,490$ \\
    OQMD\footnote{Version 1.6, downloaded in May 2025.} & $2\,732$  & $1\,731$ & $1\,334$& $5\,797$ \\
    Materials Project \& OQMD & $1\,245$ & $1\,065$ &  $1\,018$ & $3\,328$ \\
    \bottomrule
    \end{tabular*}
    \caption{
        Number of newly contributed unique crystal structures according to \texttt{pymatgen}'s \texttt{StructureMatcher} compared to the reference databases OQMD and Materials Project (and their union).
        We distinguish between three categories of new structures: \textbf{Subtype A:} Number of structures in the \gls{mc3d} with a composition that is not found in the reference.
        \textbf{Subtype B:} Number of structures in the \gls{mc3d} with an existing composition but a different spacegroup with respect to the reference.
        \textbf{Subtype C:} Number of structures in the \gls{mc3d} with existing composition and spacegroup combination with respect to the reference, but non-duplicate according to the \texttt{StructureMatcher}.
    }
\end{table*}

The $33\,142$ successfully optimized geometries were analyzed for uniqueness, using the same method used to determine the unique structures of all source structures imported from the \gls{cod}, \gls{icsd} and \gls{mpds}.
The analysis found that, after the geometry optimization, another $1\,129$ structures had become duplicates and the final set contains $32\,013$ unique optimized geometries.

Finally, we compared this set of unique structures with two other computational materials databases: the Materials Project\ccite{Jain:2013} and the OQMD\ccite{Kirklin:2015}.
These databases mostly use the \gls{icsd} as the source of input crystal structures, with the Materials Project recently also including in part structures from the \gls{mpds}, but none have considered structures from the \gls{cod} yet.
As already mentioned, another key difference with this work is that instead of \gls{qe} as the main DFT code, the Materials Project and the OQMD typically use VASP\ccite{Kresse:1996,Kresse:1999}.
Our comparison shows that \gls{mc3d} contains $3\,328$ novel structures.
Tab.~1 gives a detailed overview of the number of structures that have a spacegroup or composition that does not occur in the Materials Project and/or in the OQMD (see the Methods section for more details on how structures are matched across the different databases).

In summary, we have presented \gls{mc3d}, a database of relaxed geometries of three-dimensional stoichiometric inorganic structures computed using \gls{dft}.
The database is built starting from almost a million experimentally known crystal structures imported from the \gls{cod}, \gls{icsd} and \gls{mpds} databases.
The structures were parsed and filtered to yield a collection of $72\,589$ unique stoichiometric inorganic crystal structures, which we refer to as the \gls{mc3ds}.
We note that at most $69\,284$ of these are experimentally known, highlighting the small size of the space of experimentally known inorganic stoichiometric crystal structures.
The geometries of a large fraction of these structures were then optimized with different functionals (PBE and PBEsol) and numerical protocols using a SIRIUS-enabled version of \gls{qe}.
The most accurate and precise version of the database, \texttt{PBEsol-v2}, contains $32\,013$ unique structures.
These optimizations were performed using fully automated workflows implemented in AiiDA, automatically handling errors that occurred during the calculations, and preserving the full provenance of all data produced.
We demonstrate the workflow robustness by showing that they successfully complete for over $85\%$ of the structures, significantly improving upon the success rate without any automated error handling.
The resulting database of optimized geometries as well as their underlying provenance is made available on the Materials Cloud Archive\ccite{web:Huber:MCA:2025:mc3d} and can be browsed interactively on the Materials Cloud\ccite{Talirz:2020} portal at \href{https://www.materialscloud.org/mc3d}{https://www.materialscloud.org/mc3d}, including interactive access to the full provenance graph of the calculations.
The key distinguishing features of \gls{mc3d} include: 1) the inclusion of structures from several source databases, thus extending the coverage of materials; 2) its focus on experimentally known compounds, so that top-performing materials identified from a screening study of \gls{mc3d} are easily available; 3) the use of a consistent set of computational methods and input protocols to compute optimized geometries, providing a solid foundation for the creation of training sets for machine-learning applications; 4) the use of open-source software in the whole pipeline and in particular of \gls{qe} (and SIRIUS) as the \gls{dft} engine, at variance with many other existing databases\cite{Kirklin:2015,Jain:2013,Curtarolo:2012}, thus providing the possibility to cross-validate results with different \gls{dft} implementations; and 5) the preservation of the full provenance of all data produced by the workflows, improving the reliability of derivative work based on \gls{mc3d} and allowing other researchers to easily reproduce the results.
\gls{mc3d} thus constitutes a valuable resource for the materials science community complementing existing databases, and we expect it to be useful for applications including machine learning, materials discovery, and computational materials research.

\section*{Methods}
\subsection*{Importing crystallographic data}
The initial crystallographic data was imported from three external databases: the \gls{cod}\ccite{web:cod}, the \gls{icsd}\ccite{web:icsd}, and the \gls{mpds}\ccite{web:mpds}.
All three databases provide an \gls{api} to query and download crystal structures in the \gls{cif}\ccite{Hall:1991} format, although for the \gls{icsd} the provided MySQL database was accessed directly, without use of the \gls{api}.
AiiDA provides an interface, the \code{DbImporter} class, to implement functionality that allows importing structure from an external database.
An implementation for the \gls{cod} and \gls{icsd} already existed in the \code{aiida-core}\ccite{web:aiida-core} package, and the implementation for the \gls{mpds} was contributed for this work.
A \gls{cli} was developed to download crystal structures through these importers and store them in an AiiDA\ccite{Huber:SciData:2020} database with associated metadata, such as the \gls{url} and unique identifier used by the source database.
The code is made available through the \code{aiida-codtools}\ccite{web:aiida-codtools} Python package v3.1.0.

The \gls{mpds} reports all of their structures as experimentally observed, and the experimental temperature and pressure condition data was retrieved from the \code{condition} field of the API responses.
For the \gls{cod}, the OPTIMADE compliant API was used, and structures were flagged as theoretical if \code{\_cod\_method} field equaled \code{theoretical}.
Pressure and temperature data was retrieved from the fields \code{\_cod\_cellpressure}, \code{\_cod\_diffrpressure}, \code{\_cod\_celltemp}, and \code{\_cod\_diffrtemp}.
In the case of the \gls{icsd}, the \code{standard\_remarks} table in the 2020 MySQL version was parsed.
Remarks "THE", "ZTHE", "ABIN", "DFT" were flagged as theoretical, pressure information was parsed from the "PRE", "ZPRE" fields, and temperature was parsed from the "TEM", "ZTEM" fields.

Using these criteria, out of the $72\,589$ structures in \gls{mc3ds} we flag $3\,305$ structures as theoretical, leaving out $69\,284$ structures that might be experimentally known.
We note that $2\,214$ of these have no flag, as e.g. the 2020 version of the \gls{icsd} database does not report anymore some of the structures originally imported from \gls{icsd} version 2017.2; these might also need to be flagged as deprecated in the future, as this might point to structures that have been retracted or corrected in later versions of the database.
Because of the nature of the three source databases, we expect the vast majority of these structures to be experimentally known, even if we cannot exclude that some of them might be theoretical structures not correctly flagged as such in the source databases.

\subsection*{Cleaning and parsing crystallographic data}
The imported \gls{cif} files were cleaned using the \code{cod-tools} package v2.1\ccite{Grazulis:2015,Merkys:2016}.
The \code{cif-filter} and \code{cif-select} scripts were used to correct any incorrect syntax and remove unnecessary tags, respectively.
The \code{cif-filter} script was run with the flags \code{--fix-syntax-errors},  \code{--use-c-parser} and \code{--use-datablocks-without-coordinates}.
The \code{cif-select} script was run with the flags \code{--canonicalize-tag-names}, \code{--dont-treat-dots-as-underscores},  \code{--invert}, \code{--use-c-parser}, and \code{--tags=`\_publ\_author\_name,\_citation\_journal\_abbrev'}.
The cleaned \gls{cif} files were subsequently parsed using the \code{pymatgen} library v2018.12.12\ccite{Ong:2013} to extract the crystal structure defined through the cell basis vectors and atomic positions.
Here a tolerance of $5\cdot10^{-4}$ was employed in fractional coordinate distances to determine overlapping sites.
The parsed crystal structure was then normalized and converted to primitive cell using \code{SeeK-path} v1.8.1\ccite{Hinuma:2017}, using a symmetry precision \texttt{symprec} of $5\cdot10^{-3}$ for the underlying \code{spglib} symmetry-detection code\ccite{Atsushi:2018}.
The entire cleaning, parsing, and normalizing process is implemented in a single AiiDA workflow, called the \code{CifCleanWorkChain}, which is available from the \code{aiida-codtools}\ccite{web:aiida-codtools} Python package.
Fig.~7 shows the provenance graph produced by the execution of a \code{CifCleanWorkChain}.

After importing the \gls{cif} files with the \code{CifCleanWorkChain}, we noticed a significant number of parsed structures had a reduced formula that is inconsistent with the formula reported in the original \gls{cif}.
To address this, a post-processing step was executed that compares the chemical formula in the \gls{cif} with the chemical formula of the parsed structure.
Structures with inconsistent formulae were flagged in the database, and not considered for the next step in filtering procedure described in Fig.~1 of the Results section.
Tab.~2 gives a complete overview of the cleaning results for all structures imported from the three external databases.

\begin{figure*}
    \centering
    \includegraphics[width=0.95\linewidth]{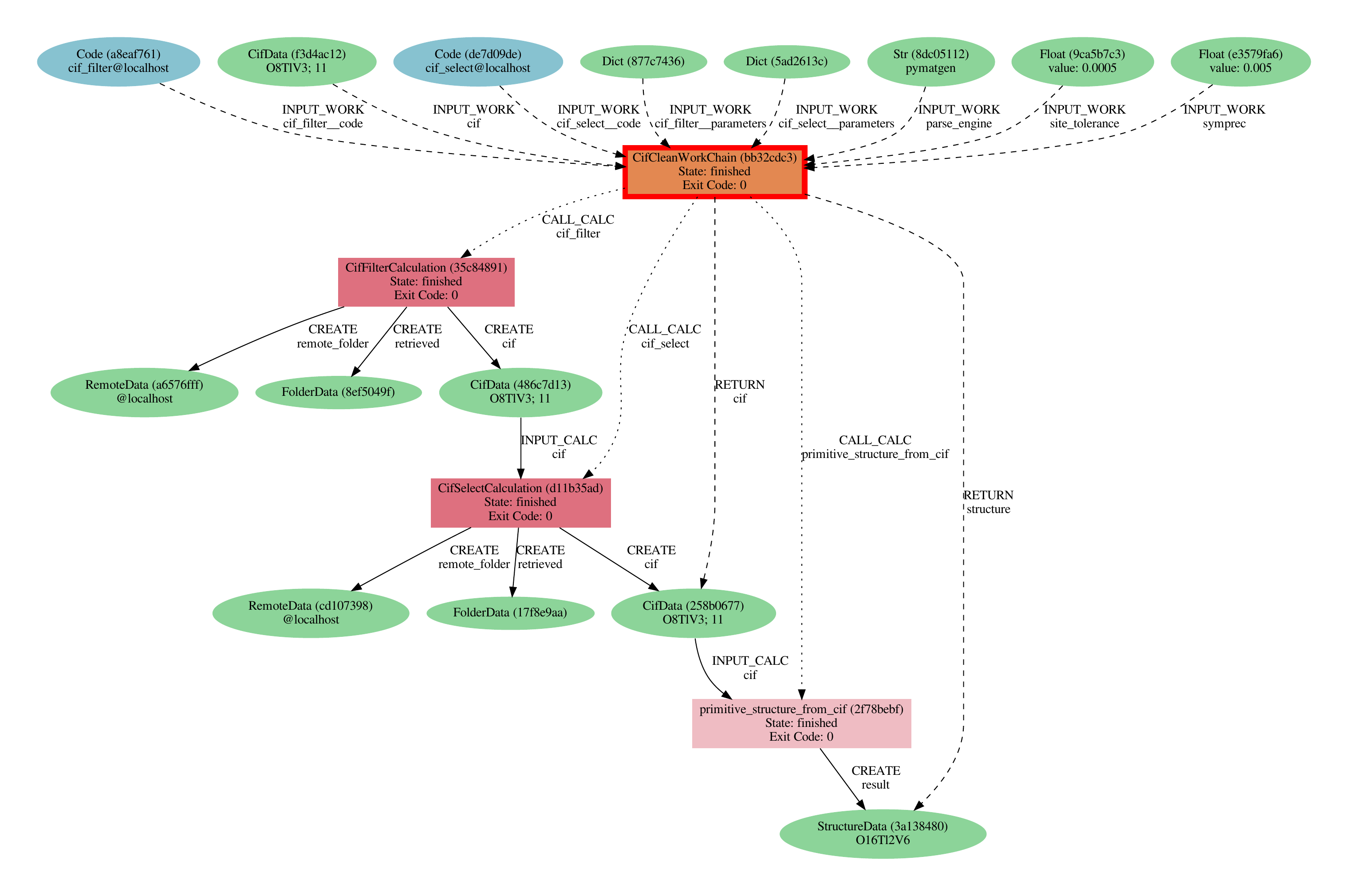}
    \caption{
        Provenance graph of an execution of the \code{CifCleanWorkChain}.
        It takes a \gls{cif} file, represented by a \code{CifData} node, as input, and uses the \code{CifFilter} and \code{CifSelect} code plugins to correct and clean the file.
        The cleaned \code{CifData} is then parsed with \code{pymatgen} to extract the structure definition (cell vectors and atom coordinates), which is then normalized and converted to primitive cell with \code{SeeK-path} into a \code{StructureData}.
    }
\end{figure*}

\begin{table*}[]
    \begin{tabular*}{\textwidth}{l@{\extracolsep{\fill}} cccr}
    \toprule
        \multirow{2}{*}{Results of the \gls{cif} cleaning and parsing} & \multicolumn{4}{c}{Occurrences} \\
    \cline{2-5}\\ &    COD &   ICSD &   MPDS & Total \\
    \midrule
        \textbf{Successfully cleaned, parsed and normalized the crystal structure }                               & \textbf{347802} & \textbf{163421} & \textbf{211942} & \textbf{723165}\\
    \arrayrulecolor{black!30}\midrule
        \textbf{Problematic \glspl{cif}} & & & \\
        \phantom{aaa}\textbf{Different composition}&  \textbf{36358} &   \textbf{7809} &  \textbf{42819} & \textbf{86986} \\[0.25ex]
        \phantom{aaaaaa}Inconsistent formula: different compositions                                                      &  36358 &   7809 &  42819 & 86986 \\
        \phantom{aaa}\textbf{Missing elements} &\textbf{9743} & \textbf{16590}& \textbf{21812} &\textbf{48145} \\[0.5ex]
        \phantom{aaaaaa}Inconsistent formula: missing hydrogen                                                            &  7426  &  15249 &  18338 & 41013 \\
        \phantom{aaaaaa}Inconsistent formula: missing lithium                                                             &    42  &     77 &    133 & 252 \\
        \phantom{aaaaaa}Inconsistent formula: missing other                                                               &  2275  &   1264 &   3341 & 6880 \\[1ex]
        \phantom{aaa}\textbf{Failed to parse} & \textbf{8314}& \textbf{5978}& \textbf{28060}& \textbf{42352} \\[0.5ex]
        \phantom{aaaaaa}The \gls{cif} had invalid syntax that could not be corrected                                     &      0 &     23 &      0 & 23\\
        \phantom{aaaaaa}The cleaned \gls{cif} defines no atomic sites                                                    &   1037 &      0 &  20645 & 21682\\
        \phantom{aaaaaa}The cleaned \gls{cif} defines sites with invalid atomic occupancies                              &   3136 &   3141 &   7415 & 13692\\
        \phantom{aaaaaa}The cleaned \gls{cif} contains sites with unknown species                                        &   1725 &   2814 &      0 & 4539\\
        \phantom{aaaaaa}\makecell[l]{The cleaned \gls{cif} defines sites with attached hydrogens with\\incomplete positional data}&   2416 &      0 &      0 & 2416\\[1ex]
        \phantom{aaa}\textbf{Symmetry issues} & \textbf{117} & \textbf{73} & \textbf{372} & \textbf{562} \\[0.5ex]
        \phantom{aaaaaa}Failed to determine the primitive structure                                                      &     34 &     70 &      8 & 112 \\
        \phantom{aaaaaa}Detected inconsistent symmetry operations                                                        &     83 &      3 &    364 & 450 \\[1ex]
    \arrayrulecolor{black!30}\midrule
        \textbf{Total}                                                                                             & \textbf{402334} & \textbf{193871} & \textbf{305005} & \textbf{901210}\\
    \arrayrulecolor{black}\bottomrule
    \end{tabular*}
    \caption{
        The results of the \code{CifCleanWorkChain} workflows and post-processing formula check.
        Each row is a potential outcome of the process, where the first column describes the outcome and the remaining columns give the number of occurrences for \glspl{cif} imported from the \gls{cod}, \gls{icsd} and \gls{mpds}, respectively.
        The "Different composition" and "Missing elements" outcomes are a result of the formula check, the other issues are direct failure modes (exit codes) of the \code{CifCleanWorkChain}.}
        The last row shows the total number of files that have been imported and analyzed with the \code{CifCleanWorkChain}.
        The bold rows indicate the corresponding (aggregated) category in the Sankey diagram in Fig.~1.
\end{table*}

\subsection*{Structure uniqueness analysis}
The structures that were successfully parsed by the \code{CifCleanWorkChain} contain many duplicates.
To determine the set of structures that are unique across all three databases, first the unique set of structures within each database was determined.
The procedure first maps all the structures of a database on their chemical formula in the reduced Hill notation\ccite{Hill:1900}.
One then only has to search for duplicates within each set of structures that share the same chemical formula, drastically reducing the computational complexity of the problem.

The second step is to separate the structures with the same formula by space group.
This is mainly to avoid cases where the structure similarity algorithm is unable to properly differentiate distinct phases with very similar unit cells.
The space group is determined using the \texttt{spglib} package with a \texttt{symprec} of 0.005.
Using this strict approach, we give preference to cases where we consider two structures to be different that, in fact, should be flagged as similar.
This is acceptable, since it is very likely that such cases will be detected as duplicates after the geometry has been optimized in the next step.
As a consequence, we will be performing a geometry optimization for more structures than strictly necessary, but we will be less likely to discard distinct but similar crystal phases that should be considered independently.

For all structures that have the same reduced formula and space group, the similarity was analyzed using the \code{StructureMatcher} utility of the \code{pymatgen}\ccite{Ong:2013} package (v2023.5.10), with a fractional length tolerance of $0.2$, a site tolerance of $0.3$ and an angle tolerance of $5$ degrees.
The options to convert to primitive cell or to attempt to match a supercell were disabled, since all structures had already been normalized using \texttt{spglib} in the previous step of the pipeline.
Finally, within each family of duplicate structures, a single representative structure was selected as the prototype.
Preference was given to structures originating from databases with more permissive usage terms, i.e. we selected from the \gls{cod} first, followed by \gls{icsd} and \gls{mpds}.

\subsection*{Removing molecular crystals}
The successfully parsed structures were filtered a last time to remove any molecular crystal structures, as they are beyond the scope of this work.
A well-defined heuristics-based algorithm to distinguish organic from inorganic crystal structures is complex to implement.
However, we note that from all three source database, only the \gls{cod} contains molecular crystal structures, and that most molecular crystal structures contain hydrogen.
The filtering was therefore performed by removing all structures originating from the \gls{cod} that are not also available in another source database and that contain hydrogen.
This method does not guarantee catching all molecular structures and it potentially also removes some inorganic crystal structures, e.g., metal hydrides, but this can be remedied in future versions of \gls{mc3d} by including again a selection of the structures that have been discarded here.
In this final filtering step, $208\,509$ structures originating from the \gls{cod} were removed.

\subsection*{Geometry optimization}
After all cleaning and filtering steps, a collection of $72\,589$ experimental unique inorganic crystal structures remains, which is referred to as the \gls{mc3ds}.
The structures of the \gls{mc3ds} form the starting point of the next part of the work that seeks to computationally optimize their geometries.

The geometry optimization is performed in several steps, implemented as an AiiDA workflow, as shown in Fig.~8.
The first step performs an initial geometry optimization with looser convergence parameters.
This helps with two aspects: (1) it is less costly to obtain a reasonable first ground state geometry, improving the efficiency and (2) the less strict parameters makes it easier to converge for geometries far removed from the ground state, improving robustness.
If the initial geometry optimization completes successfully, the optimization workflow is run again but this time with the convergence parameters as determined by the input protocol.

Both the initial and full optimization steps are instances of the same workflow, run with different input parameters.
Each workflow starts with generating the \textit{k}-point mesh used to sample the Brillouin zone based on the input geometry, such that the $k$-point density in reciprocal space matches the minimum required density specified by the input protocol.
The geometry is then optimized using \gls{qe}'s \texttt{pw.x} code\ccite{Giannozzi:2009,Giannozzi:2017} employing the SIRIUS library\cite{web:sirius} to allow running efficiently on \gls{gpu} compute nodes of different architectures.
The workflow automatically parses the output of the calculation and, in case of an error or a failure to converge, the input parameters are adjusted and restarted up to a maximum of 5 restarts.
The logic for this automated workflow is described in the ``Automated error handling'' section of the Methods.
The final optimization step is repeated until the following two conditions are met:
\begin{enumerate}
    \item The stresses in the final \gls{scf} performed by \texttt{pw.x} are below the selected convergence threshold:
    During the geometry optimization, the basis sets used (i.e. the list of reciprocal-space $G$ vectors), that depend not only on the energy cutoffs but also on the crystal geometry, are not updated at each step.
    The final \gls{scf} step that is performed at the end of the optimization recomputes the basis sets on the optimized unit cell.
    Large stresses in the final \gls{scf} indicate that the optimized geometry differs significantly from the initial geometry and the static basis sets used towards the end of the optimization were no longer converged enough.
    A restart of the optimization at the latest geometry is therefore performed.
    \item The $k$-point mesh corresponds to a lower density than the one dictated by the protocol:
    Since the unit cell can change during the optimization, it is possible that, towards the end of the optimization cycle, the initial $k$-point mesh no longer satisfies the minimum required $k$-point density specified by the input protocol.
    In this case, a new $k$-point mesh is generated and another geometry optimization is performed.
\end{enumerate}
\begin{figure}
    \centering
    \includegraphics[width=8cm]{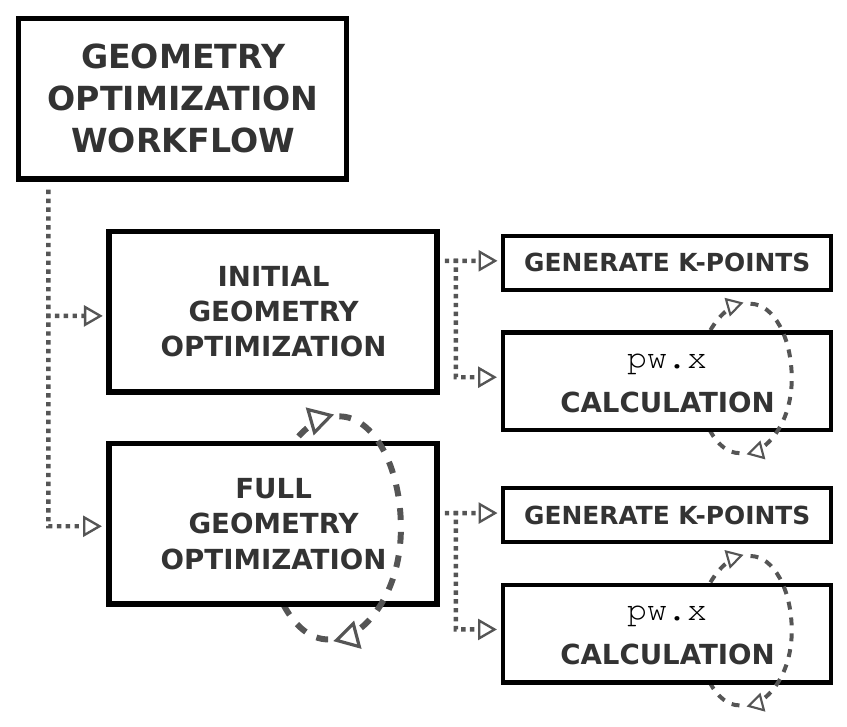}
    \caption{
        Schematic diagram of the steps of the geometry optimization workflow, as described in the text.
        A dashed circular arrow around a block indicates the step can be repeated.
    }
\end{figure}

\subsection*{Input parameter protocol}
The geometry optimization workflow takes a number of input parameters that control the precision of the calculations performed.
For the \texttt{PBEsol-v2} version of the \gls{mc3d} (see next subsection for the the \texttt{PBE-v1} and \texttt{PBEsol-v1} versions), pseudopotentials were taken from the \gls{sssp} PBEsol\ccite{Perdew:1996,Perdew:2008} efficiency v1.3\ccite{Prandini:2018,Lejaeghere:2016} library, which collects pseudopotentials from a number of libraries\ccite{Garrity:2014,Schlipf:2015,Willand:2013,DalCorso:2014,Topsakal:2014,vanSetten:2018,Kucukbenli:2014}.
\gls{sssp} provides a set of rigorously tested values for the recommended wavefunction and charge density energy cutoffs for each pseudopotential.
For each structure, the highest value among those recommended for the elements in the composition of the material was selected.
It should be noted here that recent tests versus an all-electron benchmark\ccite{Bosoni2024} have found a suboptimal precision for the \ce{Hg} and \ce{Au} pseudopotentials provided by the \gls{sssp} v1.3 efficiency.
Structures containing these elements are currently being recomputed with improved pseudopotentials and cutoffs, which will be added to the online database at a later point in time.

The magnetic and electronic properties cannot be reliably determined by just inspecting the geometry and composition of a crystal structure.
Therefore, by default, all structures are considered to be magnetic and metallic in behavior in the initial geometry optimization.
All calculations in the geometry optimization workflow are spin-polarized and are performed with smeared electronic occupations using the Marzari--Vanderbilt\ccite{Marzari:1999} cold-smearing method with a broadening of $0.02$ Ry ($\approx 2.72\cdot 10^{-1}$ eV).
Each calculation is initialized in a high-spin ferromagnetic configuration, where elements with partially occupied $d$ or $f$ orbitals are assigned a magnetic moment of 5 $\mu_B$ or 7 $\mu_B$, respectively.
For all other elements, the electrons are initialized to have a $10\%$ surplus in the spin-up channel.
To avoid an erroneous magnetic configuration being passed from the initial geometry optimization to the full one, magnetic moments are reinitialized in the second part of the workflow.

The Brillouin zone is sampled at $k$-points that are defined by a Monkhorst--Pack\ccite{Monkhorst:1976} mesh including the $\Gamma$-point, where the distance between $k$-points in each reciprocal-space direction is at most $0.15~\mbox{\AA}^{-1}$ (i.e., the algorithm chooses the smallest $k$-point mesh with at least the density required by the specified $k$-point distance).
These values correspond to the extensively tested ``balanced'' protocol described in detail by Nascimento \textit{et al.}~\cite{Nascimento:2025}.

The threshold for electronic convergence in the \gls{scf} calculation is set to $0.2\cdot10^{-9}$ Ry ($\approx2.72\cdot10^{-9}$ eV) per atom.
Geometry optimizations were stopped when the energy difference between two ionic steps was smaller than $10^{-5}$ Ry ($\approx1.36\cdot10^{-4}$ eV) per atom, all forces on all atoms were smaller than $10^{-4}~\text{Ry}/\text{Bohr}$\,($\approx 2.57 \times 10^{-3}~\text{eV}/\text{\AA}$) and cell stress was smaller than $0.5$ kbar ($0.05$ GPa).

\subsection*{MC3D versioning}
The structures of the \gls{mc3ds} have been optimized for three different versions of the geometry optimization workflows and its associated input parameter protocol.
The resulting versions of the \gls{mc3d} are named \texttt{PBE-v1}, \texttt{PBEsol-v1} and \texttt{PBEsol-v2}.
The results discussed in this article pertain to the most recent version \texttt{PBEsol-v2}.
Tab.~3 provides an overview of the differences between the three different versions.

\begin{table*}[ht]
    \centering
    \begin{tabular*}{\linewidth}{l@{\extracolsep{\fill}} c c c}
    \toprule
                                           & \texttt{PBE-v1} & \texttt{PBEsol-v1} & \texttt{PBEsol-v2}  \\
    \addlinespace
    Marzari--Vanderbilt Smearing [Ry]                          & 0.01 & 0.01    & 0.02       \\
    Pseudopotential library                & SSSP v1.2 PBE efficiency & SSSP v1.2 PBEsol efficiency & SSSP v1.3 PBEsol efficiency \\
    Geometry optimization workflow version   & v4 & v4        & v5         \\
    \bottomrule
    \end{tabular*}
    \caption{
        Differences in geometry optimization workflow and the input parameter protocol for the three different versions of the \gls{mc3d}.
        The input protocols only differ in terms of the cold smearing value used for the BZ sampling and the version of the \gls{sssp} library.
        The version of the geometry optimization workflow corresponds to the major version of the \texttt{aiida-quantumespresso}\cite{web:aiida-quantumespresso} plugin package that publishes the workflow.
    }
\end{table*}

The geometry optimization workflow, which for \texttt{PBE-v1} and \texttt{PBEsol-v1} used the workflows from \texttt{aiida-quantumespresso} version 4, was changed for \texttt{PBEsol-v2} to the workflows of \texttt{aiida-quantumespresso} version 5, with the main differences being:

\begin{itemize}
    \item for the \gls{mc3d} \texttt{v1} databases, the workflow ran a dedicated initial calculation of the electronic charge density to determine the electronic and magnetic character of the structure. This was dropped in \texttt{PBEsol-v2} in favor of reinitializing the magnetic configuration for each geometry optimization;
    \item for the \gls{mc3d} \texttt{v1} databases, the workflow did not perform an initial geometry optimization with looser convergence parameters;
    \item for the \gls{mc3d} \texttt{v1} databases, the workflow only considered the volume difference between sequential geometry optimization runs in the convergence criteria;
\end{itemize}

As mentioned before, each version of the workflow and input protocol has been applied to (a subset of) the structures of the \gls{mc3ds}, which results in different optimized versions of these source structures.
A naming convention was developed and adopted to unambiguously identify source structures of the \gls{mc3ds} and their optimized variants of the \gls{mc3d}.
All structures in the \gls{mc3d} are given a unique identifier of the form \texttt{mc3d-<source\_id>}.
Optimized geometries derived from \gls{mc3ds} structures are given a unique identifier of the form \texttt{mc3d-<source\_id>/<methodology>-<version>}.
Here the \texttt{<methodology>} and \texttt{<version>} refer to the name of the main distinguishing methodology used in the optimization workflow and the iteration, respectively.
These two concepts are separated because a change in methodology does not necessarily mean that its results supersede those of a different version.
However, for a given methodology, multiple versions of the workflow and input protocol may be run, which would in principle be improvements and the resulting optimized geometries should replace their predecessors.

\subsection*{Automated error handling}
The geometry optimization workflow consists largely of runs of the \code{pw.x} simulation code.
As presented in the Results section, these calculations have various ways of failing that need to be dealt with automatically by the workflow as much as possible for the workflow to be scalable to tens of thousands of structures.

Therefore, a logical workflow was developed to wrap the calculation in a loop and run it until it completes successfully, whose logic is schematically represented in Fig.~9.
Each time a calculation fails, predefined error handlers are called that inspect the results of the failed calculation and determine whether to abort the workflow or to perform an operation, such as changing selected input parameters, before restarting the calculation.
The workflow automatically aborts if two calculations fail consecutively without the intervention of an error handler, or the number of iterations exceeds a predefined maximum to prevent the workflow from running indefinitely.

\begin{figure*}[]
    \centering
    \includegraphics[width=16cm]{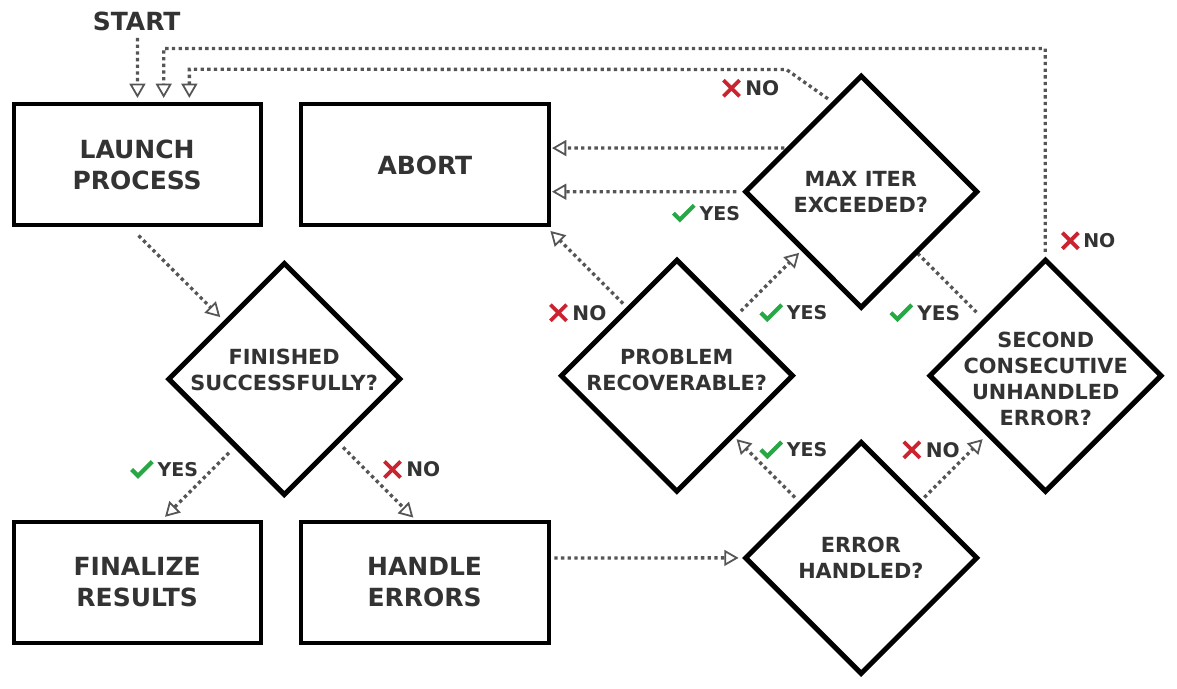}
    \caption{
        Flow diagram of the logic implemented by the error handling workflow.
        The workflow enters the main loop and starts by running the first iteration of the calculation.
        In the next step, once the calculation has terminated, in case of success the results are reported and the workflow is terminated.
        Otherwise, in the case of an error, the registered error handlers are called in order.
        If the error is handled, the calculation is restarted as long as the maximum number of iterations has not been exceeded, otherwise the workflow is aborted.
        If two consecutive calculations fail without the intervention of an error handler, the workflow is also aborted.
    }
\end{figure*}

Although specifically developed for this work, the logic of this workflow is generic and can be used in combination with any calculation.
It has since been contributed to the AiiDA workflow management system that, as of v1.1, provides it as an integrated component called the \texttt{BaseRestartWorkChain} (see AiiDA documentation\cite{web:aiida-docs-baserestart}).
There are now many AiiDA plugin packages that provide implementations of the \texttt{BaseRestartWorkChain} for a variety of codes.

\subsection*{Structure matching and comparison}
In order to determine the amount of newly added unique crystal structures compared to the \gls{mp} and \gls{oqmd}, the same parameters as described in the structure uniqueness analysis section have been adopted to initialize the \texttt{StructureMatcher}, using version \texttt{2025.5.2} of \texttt{pymatgen}.
The only difference with the structure uniqueness analysis approach was that all cells were converted to primitive cell.
This was disabled for the uniqueness analysis of the \gls{mc3ds} itself as that was already done that in a separate step before the analysis.
However, this is not necessarily the case for the structures of the \gls{mp} and \gls{oqmd}, but even the structures of the \gls{mc3ds} may have changed significantly after the geometry optimization.

To compare structures of the \gls{mc3d}, \gls{mp} and \gls{oqmd}, they were first grouped according to their spacegroup and reduced chemical formula.
Afterwards, each structure in the \gls{mc3d} was compared against all the reference structures in the corresponding subgroup to identify whether any match is detected based on the \texttt{StructureMatcher}.
To estimate the uncertainty of this analysis, since the assignment of the experimental reference to a certain structure might be handled differently in other databases, a matching was also performed on structures that refer to the same \texttt{icsd-id} (the internal identifier of structures in the \gls{icsd}, which is the source of the majority of structures in the \gls{mp} and \gls{oqmd}).
For $657$ ($2.0\%$) out of $32\,164$ common \texttt{icsd-ids} (including also tags that were filtered out as duplicates), the final structures reported in the \gls{mc3d} and \gls{mp} do not match.
However, \gls{mp} also considers the final structure to determine the reference, whereas we assign the reference database tag based on the initial structures.
In 295 cases, the final \gls{mc3d} structure does not match the initial structure, due to significant changes during geometry optimization, whereas the \gls{mp} structure does.
Further differences might be related to different XC-functionals, e.g., the PBE one adopted in the \gls{mp} (the r$^2$SCAN calculations are based on PBEsol geometry optimizations).
This high agreement justifies the application of the \texttt{StructureMatcher} approach to estimate the amount of structures in the \gls{mc3d} that are not found in the \gls{mp} and \gls{oqmd}.

\subsection*{Web platform}
Fig.~10 shows a schematic overview of the architecture of the web platform developed for the \gls{mc3d}.
The data of the \texttt{PBE-v1}, \texttt{PBEsol-v1}, and \texttt{PBEsol-v2} versions of the \gls{mc3d} are uploaded as AiiDA archive files and made publicly available on the Materials Cloud Archive.
This data is imported into an AiiDA instance running on a Materials Cloud server, and served via the AiiDA REST API, which can be explored through the Explore section of the web application.
However, while this API is powerful and gives access to the full AiiDA data, it does not provide an intuitive interface to easily find and extract relevant data for non-experts and it does not allow to serve any additional metadata that is not integral part of the AiiDA provenance graph, such as the X-ray diffraction (XRD) data that was calculated as a post-processing step without AiiDA.
Therefore, a metadata pipeline was developed which transforms the content of the \gls{mc3d} into a reduced format that just contains the optimized geometries with their relevant properties and links to the nodes in the AiiDA provenance graph that computed those properties, as well any additional metadata and derived data (such as the XRD data).
This metadata is stored in a MongoDB database which is then served via the Materials Cloud REST API, which is in turn consumed by the Discover section of the web application.

In addition to the MC3D explore and discover frontends and their related APIs, we also deploy an OPTIMADE \cite{Andersen:2021} compliant API to access the MC3D data. This makes it possible query the MC3D data in a standard manner, and explore it in any of the OPTIMADE-compliant clients available (such as the Materials Cloud OPTIMADE-client).

\begin{figure*}
    \centering
    \includegraphics[width=16cm]{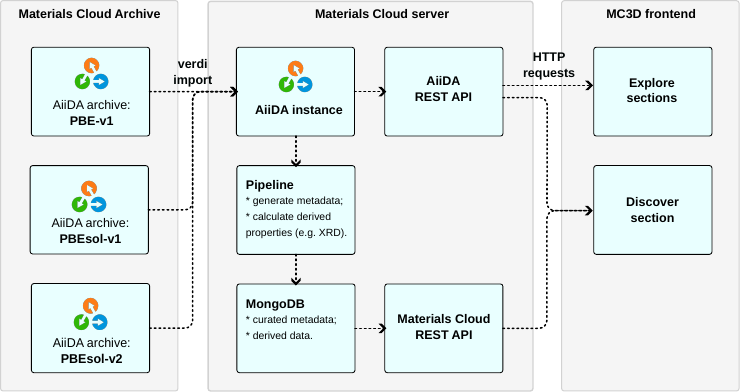}
    \caption{
        Schematic overview of the web platform developed for the \gls{mc3d}.
        The data of the different versions of the \gls{mc3d} are uploaded as AiiDA archive files to the Materials Cloud Archive.
        A Materials Cloud server serves this raw data as well as a reduced set of metadata over two REST APIs, which is consumed by the web application that provides the graphic user interface.
    }
    \label{fig:full-web-platform}
\end{figure*}

\section*{Data availability}
The databases of geometrically optimized structures are made available as AiiDA archive files on the Materials Cloud Archive\ccite{web:Huber:MCA:2025:mc3d}, together with Python scripts to extract the data from the archives into curated JSON files and then generate the figures provided in this paper.
The content of the databases can also be interactively visualized on the dedicated Materials Cloud Discover section (\href{https://www.materialscloud.org/mc3d}{https://www.materialscloud.org/mc3d}).
For structures originating from the open \gls{cod} database, the AiiDA archive files contain the complete provenance of all calculations and data, from the initial import of the \gls{cif} file from the source database to the final \gls{dft} simulations.
However, for structures originating from the commercial \gls{icsd} and \gls{mpds} datasets, we had to remove the original CIF files and the first part of the AiiDA provenance graph in order to avoid redistributing the original crystal structures, as required to comply with the licenses of these databases. Nevertheless, metadata needed to identify the original source structures, such as the source database name and structure ID, are preserved and can be accessed through the AiiDA archive files in the ``extras'' of the corresponding structures, or via the Materials Cloud frontend.

\section*{Code availability}
The source code to import, parse and clean the \glspl{cif} from the COD, ICSD and MPDS are bundled in the \texttt{aiida-codtools} package\cite{web:aiida-codtools}, which is made available under the MIT open-source license on GitHub at \href{https://github.com/aiidateam/aiida-codtools}{https://github.com/aiidateam/aiida-codtools}.
The automated workflows and input protocols to compute the optimized ground state electronic structure using Quantum ESPRESSO are bundled with the \texttt{aiida-quantumespresso} package\cite{web:aiida-quantumespresso}, which is made available under the MIT open-source license on GitHub at \href{https://github.com/aiidateam/aiida-quantumespresso}{https://github.com/aiidateam/aiida-quantumespresso}.
Both AiiDA plugin packages are distributed as installable packages through the Python Package Index, accessible at \href{https://pypi.org/project/aiida-codtools}{https://pypi.org/project/aiida-codtools} and \href{https://pypi.org/project/aiida-quantumespresso}{https://pypi.org/project/aiida-quantumespresso}, respectively.

\section*{Acknowledgments}
We acknowledge funding by the NCCR MARVEL, a National Centre of Competence in Research, funded by the Swiss National Science Foundation (grant number 205602).
We acknowledge support by the European Centre of Excellence MaX ``Materials design at the Exascale'' (Grant No. 824143).
NP was partially supported by a MARVEL INSPIRE Potentials Master's Fellowship.
NP and GP acknowledge financial support by the Swiss National Science Foundation (SNSF) Project Funding (grant 200021E\_206190 ``FISH4DIET'').
MM, MB and GP acknowledge financial support by the SwissTwins project, funded by the Swiss State Secretariat for Education, Research and Innovation (SERI).
GP acknowledges financial support by the Open Research Data Program of the ETH Board (project ``PREMISE'': Open and Reproducible Materials Science Research).

We acknowledge the EuroHPC Joint Undertaking for awarding this project access to the EuroHPC supercomputer LUMI, hosted by CSC (Finland) and the LUMI consortium through a EuroHPC Extreme Scale Access call (project 465000416) and pilot access to both the LUMI-C and LUMI-G machines under projects 465000028 and 465000170, respectively.
This work also was supported by a grant from the Swiss National Supercomputing Centre (CSCS) on Piz Daint under project IDs s836 and s854, and on the Swiss share of the LUMI system under project ID 465000106.
We acknowledge PRACE for awarding us access to Piz Daint at the Swiss National Supercomputing Centre (CSCS), Switzerland with the project id pr110.
We further acknowledge access to Piz Daint and Alps at the Swiss National Supercomputing Centre, Switzerland under the MARVEL share with the project IDs mr0 and mr32. We acknowledge the use of GH-200 nodes at the Swiss National Supercomputing Centre (CSCS) on the Tödi supercomputer within the Alps infrastructure.
This work has been partially supported by MIAI@Grenoble Alpes, (ANR-19-P3IA-0003).

We gratefully acknowledge fruitful discussions and support by Lorenzo Bastonero, Dominik Gresch, Gabriel de Miranda Nascimento, Maria Grazia Giuffreda, Michele Kotiuga, Anton Kozhevnikov, Snehal Kumbhar, Bud Macaulay, Tiziano M\"uller, Elsa Passaro, Simon Pintarelli, Giannis Savva, Leopold Talirz, Joost VandeVondele, Aliaksandr V. Yakutovich, Jusong Yu, and the members of the AiiDA and Materials Cloud team.

\section*{Author contributions}
Conceptualization: NM, GP; Methodology: SPH, MM, MB, NH, NM, GP; Supervision: NM, GP; Software: SPH, MM, MB, TR, KE, NP, MU, GP; Writing – original draft: SPH, MM, MB, TR, KE; Writing – review \& editing: SPH, MM, MB, TR, KE, NP, NH, MU, NM, GP; Validation: SPH, MM, MB, TR, KE, GP; Investigation: SPH, MM, MB, TR, KE, NP; Project administration: NM, GP; Funding acquisition: NM, GP; Formal analysis: SPH, MM, MB, TR, KE; Data curation: SPH, MM, MB, TR, KE; Visualization: SPH, MM, MB, TR, KE, NP; Resources: SPH, MM, MB, NM, GP.

\section*{References}

\end{document}